\documentclass[aps,preprint,showpacs,amsmath,amssymb,prd,nofootinbib]{revtex4-1}
\usepackage{amssymb,graphics,graphicx,epstopdf,hyperref}

\begin{document}

\title
{\Large \bf Origins of Inert Higgs Doublets}

\author{ Thomas W. Kephart$^{1}\,$\footnote{tom.kephart@gmail.com}, and Tzu-Chiang Yuan$^{2}\,$\footnote{tcyuan@gate.sinica.edu.tw}}
\affiliation{$^{1}$Department of Physics and astronomy, Vanderbilt University, Nashville, Tennessee 37235, USA\\
$^{2}$Institute of Physics, Academia Sinica, Nangang, Taipei, Taiwan 11529\\
}

\date{\today}

\begin{abstract}
 We consider beyond the standard model embedding of inert Higgs doublet fields. We argue that inert Higgs doublets can arise naturally in grand unified theories where the necessary associated $Z_2$ symmetry can occur automatically. Several examples are discussed.

\noindent
\end{abstract}

\maketitle

\section{Introduction}

With the discovery \cite{Aad:2012tfa,Chatrchyan:2012ufa} of a Higgs-like boson at about $125$ GeV 
at the Large Hadron Collider (LHC), the standard model (SM) of particle physics comes close to
its completion in terms of particle spectrum. While many of the detailed Higgs properties, 
uncannily dictated by spontaneously symmetry breaking,
still needed to be pinned down at the LHC or perhaps by the International Linear Collider (ILC) for Higgs 
precision measurements, there are existing phenomena indicating that we must extend the SM.
Among these are the neutrino masses, dark matter (DM), and baryo-leptogenesis which might be related to
TeV scale physics. On the other hand, not a single clue for new physics signal has been found in existing LHC data.

Extensions of the scalar sector beyond the lone doublet in SM 
is quite common in the literature for various reasons. 
Perhaps the most studied are the two Higgs doublet models (2HDM) \cite{Branco:2011iw}
since a second doublet is required in the popular minimal supersymmetric standard model (MSSM) \cite{Martin:1997ns},
where, with a discrete symmetry imposed upon it, a scalar field component can play the role of dark matter 
in the inert Higgs doublet model \cite{Deshpande:1977rw,Barbieri:2006dq}.
Since the 125 GeV boson behaves very much like the SM Higgs, this indicates that maybe the SM doublet 
will play the dominant role in spontaneous electroweak symmetry breaking. In other words, if there are other Higgs multiplets present in the extended scalar sector at the TeV scale, their vacuum expectation values (VEVs) must be minuscule or even vanish. Thus an inert Higgs doublet model (IHDM), in which the second doublet has  neither a VEV nor couplings with the quarks or leptons, may be a very realistic extension of the
scalar sector of the SM. 
With the upgrade of the LHC coming this year, more data will be accumulated
that could easily reveal this exciting possibility, or put stringent constraints on this simple extension. 
For detailed studies of phenomenological constraints on IHDM, 
see for example Refs.\cite{LopezHonorez:2006gr,Arhrib:2013ela}.
In this letter we study the rationale for the presence of an inert Higgs doublet at low energy. 

The paper is organized as follows.
In section II, we discuss in general how an inert Higgs doublet embedded in grand unified theories (GUTs).
In section III, we classify all the inert Higgs doublet possibilities for low lying irreducible representations (irreps) 
of frequently studied GUT gauge groups.
This is done by constructing concrete examples using $SU(5)$, $SO(10)$ and $E_6$ as our GUT gauge groups.
In section IV, we discuss some explicit models.
In section V, some phenomenological implications are discussed.

\section{Embedding the Inert Higgs Doublet in A GUT}

 It is interesting to explore how an inert Higgs models embedded in more fundamental theories.
 Let us consider grand unification theories and show that inert Higgses and their concomitant $Z_2$ 
 symmetry can arise naturally. 
 We note that there are other means for an inert Higgs doublet embedded in a higher theory, 
 for example in a composite dark sector \cite{Carmona:2015haa} 
 or in a scale invariance extension of IHDM \cite{Plascencia:2015xwa}.

Starting with an $SU(5)$ GUT where the SM Higgs lives in a $5$ and fermion families are in $(5+\overline{10})$s, it is well known that
only scalars in the 5, 10, 15, 45 and 50 irreps couple directly to fermions
since

$$\overline{10} \times 5 = \overline{5} + \overline{45},$$
$$5 \times 5 = 10 + 15,$$
and
$$\overline{10} \times \overline{10}= 5 + 45 + 50.$$
Of these, only the 5 and 45 can contain the standard
Higgs doublet. This can be seen by considering the decomposition
$$SU(5) \rightarrow SU(3)_C\times SU(2)_L\times U(1)_Y$$
where we have
$$5   \rightarrow    (1,2)_{-3}+(3,1)_{2}$$
and where $(1,2)_{-3}$ is the standard Higgs. For convenience the $U(1)_Y$ charges are all taken to be
integers by our choice of normalization.
Similarly
$$ 45 \rightarrow    (1,2)_{-3}+(3,1)_{2}+(\overline{3},1)_{-8}+(\overline{3},2)_{7}+(3,3)_{2}+(\overline{6},1)_{2}+(8,2)_{-3}$$
while the 10, 15 and 50 contain no SM doublets, i.e., they have no $(1,2)_{-3}$ in their decomposition.

There are other low dimensional irreps containing electroweak (EW) doublets $(1,2)_{x}$, 
but we need 3 or $-3$ for the $U(1)_Y$ charge $x$ for the doublet to be of normal Higgs type. The
only other $SU(5)$ irrep
with dimension less than 200 that also contains a SM EW doublet with the right
$U(1)_Y$ charge is the 70

$$ 70 \rightarrow    (1,2)_{-3}+(3,1)_{2}+(1,4)_{-3}+(3,3)_{2}+(\overline{3},3)_{-8}+(6,2)_{7}+(8,2)_{-3}+(15,1)_{2}$$
But the 70 does not couple directly to SM fermions. If you tune the
mass parameter of the 70 in the scalar potential to be positive (also assuming proper choices of scalar quartic couplings), then it will not get a VEV. So to lowest
order the 70 contains a doublet that acts  a lot like the inert Higgs. I.e., the 70 contains a SM EW doublet of the right charge that does not get a VEV and does not couple to the SM fermions. But what is lacking is a $Z_2$ symmetry to complete the model and let the doublet in the 70 be properly identified as an inert Higgs.

There are additional issues with the 70. For instance, if we have a 24 Higgs to break
$SU(5)$ to the SM gauge group $SU(3)\times SU(2)\times U(1)$ and since
$$70\times 24    =     5+45+2(70)+280+280'+450'+480 \; ,$$
then we would get dimension 5 operators like
$$\overline{10}_F 5_F 70_H 24_H \; , $$
which could be  problematic for phenomenology. But there may also be ways around such problems.
E.g., instead of the 70 we could choose a different higher dimensional
irrep that
contains a $(1,2)_{-3}$ but has no operators coupling it to fermions up to some large mass
dimension $N$, where $N$
depends on the choice of irrep.

We could also avoid the 24 and break to $SU(3)\times SU(2)\times U(1)$ with
something else like a 75 to change things.
As we will see below, SUSY can also be used to help with some of the issues.

\section{Classification}

 It is useful to classify all
the inert Higgs possibilities for the popular GUT groups
like $SU(5)$, $SO(10)$ and $E_6$. The criteria for the inert Higgs
are the following:
\begin{itemize}
\item[(1)]
It has no VEV;
\item[(2)]
It does not couple to SM fermions;
\item[(3)]
It is odd under a $Z_2$ symmetry under which all the SM particles transform trivially.
\end{itemize}

A number of  questions arise as to the nature of inert Higgses. Could EW scalar doublets with non standard $U(1)$ charges
be of interest? Could they also play the part of an
inert  Higgs? I.e., are they close enough to being inert Higgses
that they can deliver the same or similar phenomenology?
Are there other ways the idea of inert Higgs can be generalized? We will address some of these questions below.

Examples of EW scalar doublets in $SU(5)$ with non-standard charges are:
(a) the 40 contains a doublet $(1,2)_9$;
(b) the $175'$ has a $(1,2)_{-15}$.
There are of course more examples in higher dimensional irreps.
Here we give a classification of irrep of $SU(5)$ (up to dimension 1000), irreps
of $SO(10)$ (up to dimension 4000) and irreps of $E_6$ (up to dimension 20,000) that contain SM doublets with standard hypercharge, SM doublets with non-standard hypercharge, or both.

\subsection{SU(5)}

There are 51 irreps of $SU(5)$ with dimensions less than or equal to 1000, but
there are not that many (only 12) that contain EW doublets. A systematic collection of these results is given in Table \ref{SU(5)doublets}. Note that none of these irreps contain more than one doublet.
As we see, only 6 of the 12 have the standard EW hypercharge. (These results and those given below are all easily checked using the software package LieART~\cite{Feger:2012bs}.)

\begin{table}[htdp]
\caption{$SU(5)$ irreps of dimension less than 1000 that contain $SU(3)\times SU(2)\times U(1)$ doublets.}
\begin{center}
\begin{tabular}{c|c|c|c|c}
$SU(5)$ irrep & doublet&~~~~~~~~&$SU(5)$ irrep & doublet\\
\hline\hline
5 & $(1,2)_{-3}$&&280&$(1,2)_{-3}$\\
40 & $(1,2)_{9}$&&450&$(1,2)_{9}$\\
45 & $(1,2)_{-3}$&&$450'$&$(1,2)_{-3}$\\
70 & $(1,2)_{-3}$&&480&$(1,2)_{-3}$\\
$175'$ & $(1,2)_{15}$&&$560'$&$(1,2)_{-21}$\\
210 & $(1,2)_{9}$&&$700'$&$(1,2)_{15}$\\
\end{tabular}
\end{center}
\label{SU(5)doublets}
\end{table}

\subsection{SO(10)}

For $SO(10)$ all the  fermions are in $16$s, where
$$16 \times 16 = 10 + 120 + 126 .$$
To couple to fermions,   a Higgs must be in a 10, 120 or $\overline{126}$ (both the 10 and 120 are real irreps).

Under the decomposition
$$SO(10) \rightarrow SU(5) \times U(1)  \rightarrow SU(5)$$
we have
$$10 \rightarrow 5_2+\overline{5}_{-2}\rightarrow 5 + \overline{5},$$
$$120  \rightarrow  5_2+\overline{5}_{-2}+10_{-6}+\overline{10}_{6}+45_2+\overline{45}_{-2}  
\rightarrow 5 + \overline{5} + 10 + \overline{10} + 45 + \overline{45},$$
and
$$126 \rightarrow   1_{10}+5_2+\overline{10}_{6}+15_{-6}+\overline{45}_{-2} +50_2 \rightarrow
1 + 5 + \overline{10} + 15 + \overline{45} + 50.$$
Again the SM Higgs can only be in the 5 or 45 of $SU(5)$ since only they contain
the $(1,2)_{-3}$ that couples to fermions.

To find other doublets in $SO(10)$ irreps with SM charges, we can just find those
$SU(5)$ irreps where
the SM doublets can live -- i.e., those on the list above in Table \ref{SU(5)doublets}. If we insist
on standard EW charged
doublets we look for $SO(10)$ irreps that contain 5, 45, 70, 280, $450'$, or
480 of $SU(5)$. Besides the 10, 120 and 126 they are (for $SO(10)$ irrep dimension less than 4000)
16, 144, 210 $210'$, 320, 560, 672, 720, 945, 1050, 1200, 1440, 1728, 1782, 2640, 2772, 2970, 3696 and
$3696'$. 
We note that an inert Higgs doublet embedded in the 16 of $SO(10)$ was studied previously 
in \cite{Kadastik:2009dj}.

Non standard hypercharged doublets live in the following $SO(10)$ irreps (again for dimension less than 4000):
144, 210, 320, 672, 720, 945, 1050, 1200, 1440, 1728, 2640, 2772, 2970, 3696 and $3696'$.

Note that unlike the $SU(5)$ irreps in Table~\ref{SU(5)doublets}, there are $SO(10)$ irreps on the above list containing more than one doublet. So there is a lot of overlap with many $SO(10)$ irreps containing both
doublets with standard and non-standard hypercharge. To demonstrate, here is one
detailed $SO(10) \rightarrow SU(5) \times U(1)$ example:\\
\begin{eqnarray}
3696'   &\rightarrow &   1_{10}+5_2+10_{14}+2(\overline{10})_6+15_{-6}+24_{10}+\overline{40}_6+45_2+2(\overline{45})_{-2}+2(50)_2 \nonumber\\
&+&70_2+75_{10}+\overline{105}_{-2}+126_{-10}+160_{-6} 
+175_{-6}+\overline{175}_6+\overline{175''}_{-2}+210_{-6}+280_2 \nonumber\\
&+&\overline{280}_{-2}
+\overline{315}_6+\overline{480}_{-2}+720_2 \nonumber
\end{eqnarray} 
where the 5, 45, etc. $SU(5)$ irreps  contain standard hypercharged
doublets, and the $\overline{40}$, 210, etc.  irreps of $SU(5)$ contain non-standard
hypercharged doublets. Note that it would be easy to extend our analysis to flipped $SU(5)$ 
models, or other more general flipping in $E_6$ models (For  a summary see e.g., \cite{Kephart:1989az} and references therein.), but we will resist the urge for sake of brevity.

\subsection{E$_6$}

Let us now turn to the discussion of $E_6$. By decomposing the $27_H$ through the decay chain
\begin{eqnarray}
E_6\rightarrow SO(10)    \rightarrow SU(5) \nonumber
\end{eqnarray}
we find 
\begin{eqnarray}
27\rightarrow 16+10+1     \rightarrow (10+\bar{5}+1)+(5+\bar{5}+1) \; . \nonumber
\end{eqnarray}
Consulting Table \ref{SU(5)doublets} we see that the $27_H$ contains doublets with SM EW charges in the $5$s, and none with exotic hypercharge. We have looked at the same decomposition of  all $E_6$ irreps that live entirely in the cascade with $SO(10)$ irreps of dimension less that 4,000 which in turn decompose into $SU(5)$ irreps of dimension less than 1,000. For $E_6$ this means we need all irreps of dimension less that 20,000 except for the $19,305'$.
As we have just seen, the  doublets in the $27$ all have Higgs like SM EW charges. This is also true of the $78$. All other irreps of $E_6$ on our cascade list contain both normal and exotic charged EW doublets. Let us consider just one example, the 351.

\begin{eqnarray}
351\rightarrow 10+16+\bar{16}+45+120+144 
    \rightarrow (5+\bar{5})+(10+\bar{5}+1)+(\overline{10}+5+1) \nonumber\\
    +(24+10+\overline{10}+1)
    + (5+\bar{5}+ 10+\overline{10}+ 45+\overline{45}) 
    +(5+\bar{5}+10+15+24+40+\overline{45})\nonumber
\end{eqnarray}
Consulting Table \ref{SU(5)doublets} we see that only the 40 contains a doublet with exotic hypercharge, all the other doublets from the $351$ live in 5s and 45s and have normal hypercharge. All other $E_6$ irreps under consideration contain 5s and 45s plus at least one 40 or a 210 (which also contains an exotic EW doublet) of $SU(5)$, hence we reach our conclusion that these $E_6$ irreps all contain both normal and exotic doublets.

\section{Models}

 The above classification  tells us how
general $n$-Higgs doublet models (nHDMs), and in particular  the 2HDM, can fit into a GUT scheme.
To get inert doublets without imposing a $Z_2$ directly, we have
to forbid Yukawa couplings as well. For simplicity, in this section we just focus  on $SU(5)$ models.

Since only the $5_H$ and $45_H$ of $SU(5)$ couple directly to fermions at dimension 4, the
other scalar irreps have to couple via higher dimensional operators and how that
works would depend on what else is in the model.
In this sense renormalizable Yukawa couplings for all doublets not in $5_H$ and $45_H$ are automatically  forbidden at dimension 4 by the group theory.
Then it is up to the model builder to forbid as many higher
dimensional operators as necessary to generate a good model.

Say we have a minimal $SU(5)$ model which contains the usual $5_H$ and  $24_H$
augmented by an additional $480_H$ which contains our extra standard charge Higgs doublet. 
Since
$$24\times 24     =     1+2(24)+75+126+\overline{126}+200 \; ,$$
upon taking the product $24\times 24\times 5$ we find
$$\overline{126}\times 5     =     45+105+480$$
and
$$200\times  5     =     70+450'+480$$
both of which contain a 480,
so there are at least two dimension 8 operators
$$(\overline{10}_F 5_F) (24_H 24_H 5_H \overline{480}_H) 5_H$$
that couple the $480_H$ to fermions.

If we put the additional $SU(5)$ Higgses in properly chosen  irreps of higher dimension, then
we would expect the coupling to fermions could be put off to even higher
dimensional operators. But this may no be the most attractive thing to do.
 
There is another possibility -- replace the $24_H$ with
a higher dimensional irrep (we mentioned the possibility of using a $75_H$ above). All we need is an irrep with a
neutral $SU(3)\times SU(2)\times U(1)$ singlet that can get  a VEV to
break $SU(5)$ to $SU(3)\times SU(2)\times U(1)$.
E.g., the $1000_H$ of $SU(5)$ is such an irrep. Now put the extra doublet $(1,2)_{-3}$
in a $70_H$. Since
$$\overline{70}\times 70     =     1+2(24)+75+126+\overline{126}+2(200)+1000+1024+1050'+\overline{1050'}$$
contains the $1000$ there is an operator of dimension 7
$$(\overline{10}_F 5_F) (1000_H 70_H \overline{70}_H) 5_H \; .$$
Clearly the lowest dimensional operator where the new Higgs couples depends on the choice of scalar irreps.

We already have a list of $SU(5)$ irreps that contain SM doublets
that do not couple to fermions in 5s and $\overline{10}$s.
What is missing is the $Z_2$, but we now show that we are able to get that too in a natural way.
Let  $R_H$ be an irrep containing a SM doublet that does not couple to
SM fermions. We want  $R_H$ to be odd under a $Z_2$ with the rest of
the SM particles even. The can only happen without fine tuning if there are no
terms in the most general Lagrangian that are linear or cubic in  $R_H$.
(To begin with, we consider only renormalizable non SUSY Lagrangian.)
The kinetic term for  $R_H$ is quadratic and there are no Yukawa term that include  $R_H$, so the only place where there could be trouble is  in the Higgs potential.

To be specific, let us again consider minimal $SU(5)$ with Higgses  $5_H$ and $24_H$ and
extend it by adding the irrep $R_H$. Potential problem terms in the scalar potential are of the form
$H_1 H_2 R_H$ or $(R_H )^3$ or $ \overline{R}_H\times (R_H)^2,$
where $H_i$ is $24_H$, $5_H$ or $\overline{5}_H$.
So we must require that  $R_H$ has no cubic invariant and is
not in the products of 24s and 5s and $\overline{5}$s.
We also have to avoid quartic terms that are linear or cubic in  $R_H$.

The first irrep not coupling to fermions at dimension 4 and containing a  doublet with SM hypercharge is the 70, but there is a singlet in $70\times \overline{5}\times 24$, so this term is not $Z_2$ invariant.

The next irrep with a SM doublet with standard hypercharge is the $280_H$.
It has no singlet in any of the cubic terms, including $(280)^3$ and
$\overline{280} \times (280)^2$.
But there are quartic terms cubic in $280_H$. E.g., $(280)^3\times \overline{5}$ has a singlet.
So the Lagrangian for minimal $SU(5)$ plus a $280_H$ fails to have a $Z_2$ invariance.
It is easy to check that the 480 also fails because there is a singlet in
$(24)^2 \times \overline{5} \times 480$.

We observe that if the $Z_2$ arose as an accidental symmetry it would be some similarity with the ``automatic invisible axion" where the choice of irreps left an accidental $U(1)$ that could be identified with $U(1)_A$.
Here we seek an accidental $Z_2$ that delivers an  automatic inert Higgs doublet.
It appears we need to go to quite high irreps to make this happen in non-SUSY extended $SU(5)$ model. However,
if we allow SUSY  we can easily arrive at  viable automatic inert Higgs doublet models, because if the cubic terms do not violate $Z_2$ invariance, then that is all we need, since then all the quartic terms in the Higgs potential  come from cubic terms in the superpotential and
problem quartic terms like $(280)^3\times \overline{5}$ never appear.
Hence we immediately have two automatic $Z_2$ symmetric inert Higgs doublet examples which are the MSSM extended with either a $280_H$ or a $480_H$. The $Z_2$ remains unbroken and the model delivers a DM candidate if the irrep extending the MSSM does not get a VEV.

\section{Discussions and Conclusions}

Here we will discuss some possible phenomenological implications based on our findings in the previous section,
continuing to focus on the special case of $SU(5)$.

\begin{itemize}
\item
In the $SU(5)$ examples above, if the $280_H$ (or the $480_H$) does not get a VEV, then the $Z_2$
never gets broken, so the lightest component of the associated inert Higgs doublet will be a DM candidate. 

\item
All the doublets in $SU(5)$ irreps in Table~\ref{SU(5)doublets} with dimension less than 1000 have
standard hypercharge except for a few, and these
nonstandard hypercharge doublets have high
electric charges. These are:
the 40, 210 and 450 with electric charges $(\pm 1, \pm 2)$,
the   $175'$   and $700'$ with electric charges $(\pm 2, \pm 3)$,
and the $560'$ with electric charges $(\pm 3, \pm 4)$.
All the electric charges are integers and are $(\pm n, \pm n  \pm 1)$
for the four states in one of the doublets -- call it $\Phi_n$.
So the SM Higgs doublet is a $\Phi_0$, the  $175'$  
contains a $\Phi_2$, etc. For the $n > 0$ cases the 
inert Higgses can not get a VEV without breaking electric charge, since they have no neutral
components. Also since they can not couple to fermions, they can only couple to
the EW gauge fields via the standard gauge interactions 
and the SM Higgs doublet $H$ via terms of the form
$(H^\dagger H)(\Phi_n^{\dagger} \Phi_n) \; ,$
except for a $\Phi_0$ which can also couple via
$
(H^\dagger \Phi_0)(\Phi_0^\dagger H) \; .
$
For $n > 0$ we have seen examples where a $Z_2$ symmetry can arise automatically (accidentally) so we may be able to avoid explicitly imposing extra global symmetries like $Z_2$ or $S_3$ to allow $\Phi_n$ to become a
DM candidate. The SM Higgs doublet $H$ would then be the portal connecting the DM to the visible SM sector.

\item
For $n > 0$ the inert Higgses have to appear in pairs and the lightest component would be stable. They can annihilate pair-wise 
into photons, but after freeze-out they could form neutral ``atoms" and be part of the dark matter. 
For instance the charge 2 component $\phi^{++}$ of $\Phi_2$  could bind two electrons to form a
helium-like atom (dark helium atom). The energy levels would be only slightly shifted from true helium since the nucleus would weigh a few TeV instead of 4 GeV. These particles could be easily hidden from observations. The $\phi^{--}$ component may be harder to hide since it would need to bind either to positrons, which are probably not available, or protons which would have helium like energy levels but shifted  into the X-ray spectrum. 

\item
One could also think of having all the DM in the lightest stable inert Higgs state, say $\phi^{++}$ and look for an ``apparent''
excess of helium from standard BBN predictions. (Here, to simplify the discussion, we assume a $\phi^{++}/\phi^{--}$ asymmetry.)
If the $\phi$ mass is  ~1 TeV, then a pseudo-He atom is 250 times as heavier as normal He.
From BBN we know 25\% of the baryons are in He. We have about 5 times as much energy density in DM as baryons.
So 1 TeV pseudo-He DM would contribute what appears like a 0.5\% excess in He in the Universe, which is probably close
to the detectable range. If the lightest stable inert Higgs state is say $\phi^{+3}$, then we'd get an apparent excess of pseudo-Lithium. Since they are predicted to be more rare, Li and other heavy elements would give much stronger limits than H or He, and so we could expect to get strong bounds on
the $\phi^{+n}$ masses, for $n >2$.

\item
Free stable $\phi^{\pm n}$s  could potentially be primary cosmic ray components. They would be  charged (i.e. charge $n$) heavy particles without strong interactions. They would be highly penetrating  like muon but difficult to accelerate to relativistic velocities because of  their small charge-to-mass ratio. 
However, we note that cosmic charged stable particles are usually considered to be excluded by 
cosmological arguments coupled with terrestrial searches for 
anomalously heavy water molecules \cite{Byrne:2002ri}.

\item
The renormalization group (RG) running of the hypercharge $U(1)_Y$ coupling would be faster when we include an extra  inert Higgs, so we would need to add color
thresholds to compensate in order to preserve unification.
To be more specific, first we note that by adding  particles with large hypercharges the $U(1)_Y$ 
coupling grows even faster with mass scale. 
Secondly, we also change the
$SU(2)_L$ running since we are adding EW doublets. 
For the RG trajectories to unify at the same place (say around $10^{16}$ GeV for MSSM), 
we then would need to change the $SU(3)_C$ running (and probably adjust the $SU(2)$ a bit too) 
so that it bends in the same direction as the $U(1)_Y$.
This requires particle with color charges, i.e.,  quarks or maybe exotics.
Thus finding a highly charged doublet could indicate a 4$^{\rm th}$ family at a fairly low scale 
(ignoring other problems with having a 4$^{\rm th}$ family).

\end{itemize}

A $Z_2$ discrete symmetry, unbroken at tree level in a renormalizable model, never gets broken by loop diagrams (higher order operators). All terms in the Lagrangian are even in $R_H$, and since $R_H$ gets no VEV,   all higher order operators are also even in $R_H$ and hence conserve the $Z_2$ symmetry. 

We can consider these models as
UV completion of the inert Higgs doublet -- at least up to
near the GUT scale $\sim 10^{16}$ GeV. Beyond that we need to worry about
the fact that quantum gravity effects can violate any global discrete symmetry \cite{Krauss:1988zc}. 
The way to avoid this problem is to gauge the $Z_2$, promoting it into local discrete symmetry \cite{Krauss:1988zc},
but that is beyond the scope of our present analysis.

Besides decoupling an inert Higgs $H'$ doublet from SM fermions we have seen that we can also decouple it from the SM Higgs $H$. (We call these cases `strongly inert'.) Examples include the MSSM extended by either $280_H$ or a $480_H$, both of which deliver automatic $Z_2$s.
For this reason we have been lead to broaden our definition of what we mean by an inert Higgs. We can also generalize $Z_2$ to any discrete group~\cite{Dent:2009pd,Albright:2012zt,Albright:2012bp}, either abelian or nonabelian that accommodates either one of these decouplings. One could even have multiple inert and/or strongly inert Higgses. These alternative systems will have phenomenology that differs from the standard IHDM. In particular since the $H'$ in the strongly inert case only couples to gauge bosons, the global fit results found in \cite{Arhrib:2013ela} will require modification. 
Phenomenology of the generalized inert Higgs explored in this work is quite rich and further study is deserved.

\acknowledgments
The work of TWK was supported in part by US DoE grants DE-FG05-85ER40226 and DE-SC0010504, and in part by National Science Foundation Grant No. PHYS-1066293 and by the hospitality of the Aspen Center for Physics.
The work of TCY was supported in part by the Ministry of Science and Technology (MOST), Taiwan, ROC under the Grant Nos. 101-2112-M-001-005-MY3 and 104-2112-M-001-001-MY3.

\end{document}